\documentclass[aps,twocolumn,pr*,amsfonts,showpacs,superscriptaddress,longbibliography]{revtex4-2}
\DeclareUnicodeCharacter{0393}{$\Gamma$}
\usepackage{verbatim,epsfig,amsmath,amssymb,bm,epsf,graphicx,psfrag,bbold,amsthm,amsfonts}
\usepackage[bottom]{footmisc}
\usepackage{hyperref}
\usepackage{cleveref} %should be placed only after hyperref
\usepackage{enumitem}
\usepackage{framed}
\usepackage{mathrsfs}
\usepackage{esint}
\setlist{nosep}
\usepackage{placeins}
\usepackage[all]{xy}
\usepackage{color}
\usepackage[utf8]{inputenc}
\usepackage{float}
\usepackage{natbib}
\usepackage{tikz-cd}
\usepackage{verbatim}
\usepackage{leftidx}
\usepackage{physics}
\usepackage{ulem}

\usepackage{pifont}% http://ctan.org/pkg/pifont
\usepackage{braket}
\usepackage{booktabs}

\usepackage{tikz-cd}
\usetikzlibrary{positioning}

\begin{document}
%\title{Artificial Intelligence for Quantum Matter: Finding a Needle in a Haystack}
% \title{QERNEL: Quantum Expert-Routed Neural Learner}
% \title{QERNEL: Efficient Foundational Model For Quantum Matter}

% \title{Discovering Quantum Phase Transition with Foundational AI}

\title{QERNEL: a Scalable Large Electron Model}

% \title{A Scalable Large Electron Model Discovering Quantum Phase Transitions}

\author{Khachatur Nazaryan}
\email[Email: ]{khachnaz@mit.edu}
\affiliation{Department of Physics, Massachusetts Institute of Technology, Cambridge, MA-02139,USA}

\author{Liang Fu}
\affiliation{Department of Physics, Massachusetts Institute of Technology, Cambridge, MA-02139,USA}

\date{\today}

\begin{abstract}
We introduce QERNEL, a foundational neural wavefunction that variationally solves families of parameterized many-electron Hamiltonians and captures their ground states throughout parameter space within a single model. QERNEL combines FiLM-based parameter conditioning with scale-efficient architectural elements --- mixture of experts and grouped-query attention, substantially improving expressivity at low computational cost. We apply QERNEL to interacting electrons in semiconductor moir\'e heterobilayers, training a single weight-shared model for systems of up to 150 electrons. By solving the many-electron Schr\"odinger equation conditioned on moir\'e potential depth, QERNEL captures both quantum liquid and crystal states and discovers the sharp phase transition between them, marked by abrupt changes in interaction energy and charge density. Our work establishes a foundation model for moir\'e quantum materials and  a scalable architecture toward a Large Electron Model for solids. 
\end{abstract}

\maketitle

\section{Introduction}

Representing many-body wavefunctions with neural networks has become a powerful variational method for solving interacting quantum systems in continuous space. In this framework, the neural network itself defines the variational ansatz: its weights and biases parameterize the wavefunction, and are optimized by minimizing the Monte Carlo estimate of the energy. In recent years, this approach has delivered highly accurate ground states for atoms and molecules \cite{Pfau2020Sep,PauliNet2020,vonGlehn2022Nov,Li2023JulForwardLaplacian}%,Gao2023Jul}
, the uniform electron gas \cite{Cassella2023Jan,Wilson2023Jun,Luo2023,Kim2024May,Smith2024,pescia2023message}%,sobral2024physics}
, moir\'e semiconductors \cite{Li2024,Luo2024,GeierNazaryan2025}, and fractional quantum Hall liquids \cite{Teng2024Nov,Qian2024,NazaryanGaggioli2025}. %, while related neural-wavefunction ideas have also been applied successfully to lattice models \cite{Carleo2019Jul,Moreno2022,viteritti2023transformer,Luo2023Mar,ChenJ1J22024}. 
Much of this progress has been achieved with determinant-based architectures, which strictly enforce fermionic antisymmetry. As shown recently, a linear combination of Slater determinants multiplied by symmetric functions can approximate any continuous Fermi wavefunctions to arbitrary accuracy \cite{fu2025,fu2026}. This universal representational power is realized in neural networks through the attention mechanism \cite{GeierNazaryan2025}, which builds many-body correlations between particles.

Yet despite this success, the framework still mostly uses neural network in a pointwise way: one chooses a Hamiltonian, optimizes a neural wavefunction for that instance, and repeats the optimization when the parameters change. From the perspective of numerical many-body physics this is natural, since the goal is to solve each problem as accurately as possible. From the perspective of modern AI, however, it is a limited use of neural networks. There the aim is not merely to solve one task, but to learn an underlying pattern that generalizes across a family of related tasks. This raises a natural question for quantum many-body physics: can one train a single neural wavefunction that generalizes across an entire family of Hamiltonians?

\begin{figure}
    \includegraphics[width=\linewidth]{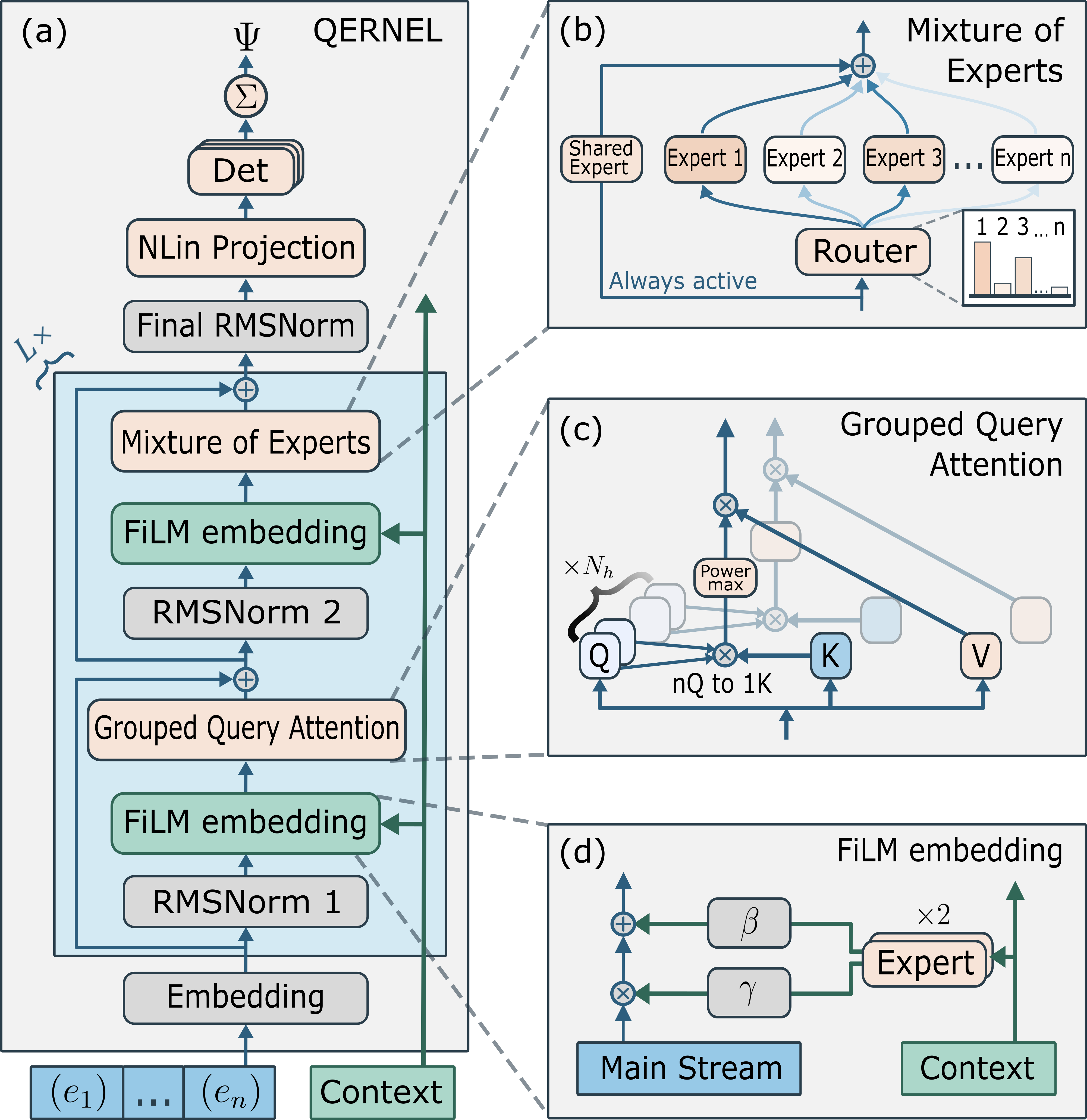}
    \caption{Architecture of QERNEL. (a) Overall conditional wavefunction pipeline: electron features are embedded into electronic streams, processed by stacked equivariant blocks, projected to generalized orbitals, and combined into determinants. The Hamiltonian context is threaded alongside the electron streams and repeatedly injected throughout the depth of the network. (b) Mixture-of-experts block with a router and an shared expert. (c) Grouped-query attention, where multiple query heads share key and value heads for improved efficiency. (d) FiLM conditioning, which injects the context into the electronic streams through channel-wise shifts and rescalings.}    \label{fig:architecture}
\end{figure}

Recent work has begun to explore this direction. A unified neural ansatz for multiple Hamiltonians in lattice systems  has been developed \cite{Rende2025FNQS}. In quantum chemistry, neural wavefunction optimization can be amortized across molecules and geometries \cite{Foster2025Orbformer}. Recently, a foundation model based on universal Fermi network \cite{fu2026} has been introduced to solve many-electron Schrodinger equation across multiple Hamiltonian parameters and particle numbers and predict ground states for unseen instances \cite{Zaklama2025AttentionFM,Zaklama2026LEM}. This ``large electron model'' points toward a paradigm shift from single-instance wavefunction solvers to foundation models that learn generalizable wavefunctions.  

As the ambition for quantum foundation models grows, a limitation becomes increasingly severe: compute. To attain accurate results often require models with parameter counts approaching or exceeding \(10^6\) even for systems of only $\sim 20$ particles \cite{GeierNazaryan2025}. %, together with substantial hardware resources \cite{GeierNazaryan2025}. 
This makes the methodology powerful, but not always accessible to the broader community. If foundational neural wavefunctions are to become practical, the architecture itself must become more efficient, more stable, and more explicitly designed for generalization.

\begin{figure}
    \includegraphics[width=0.9\linewidth]{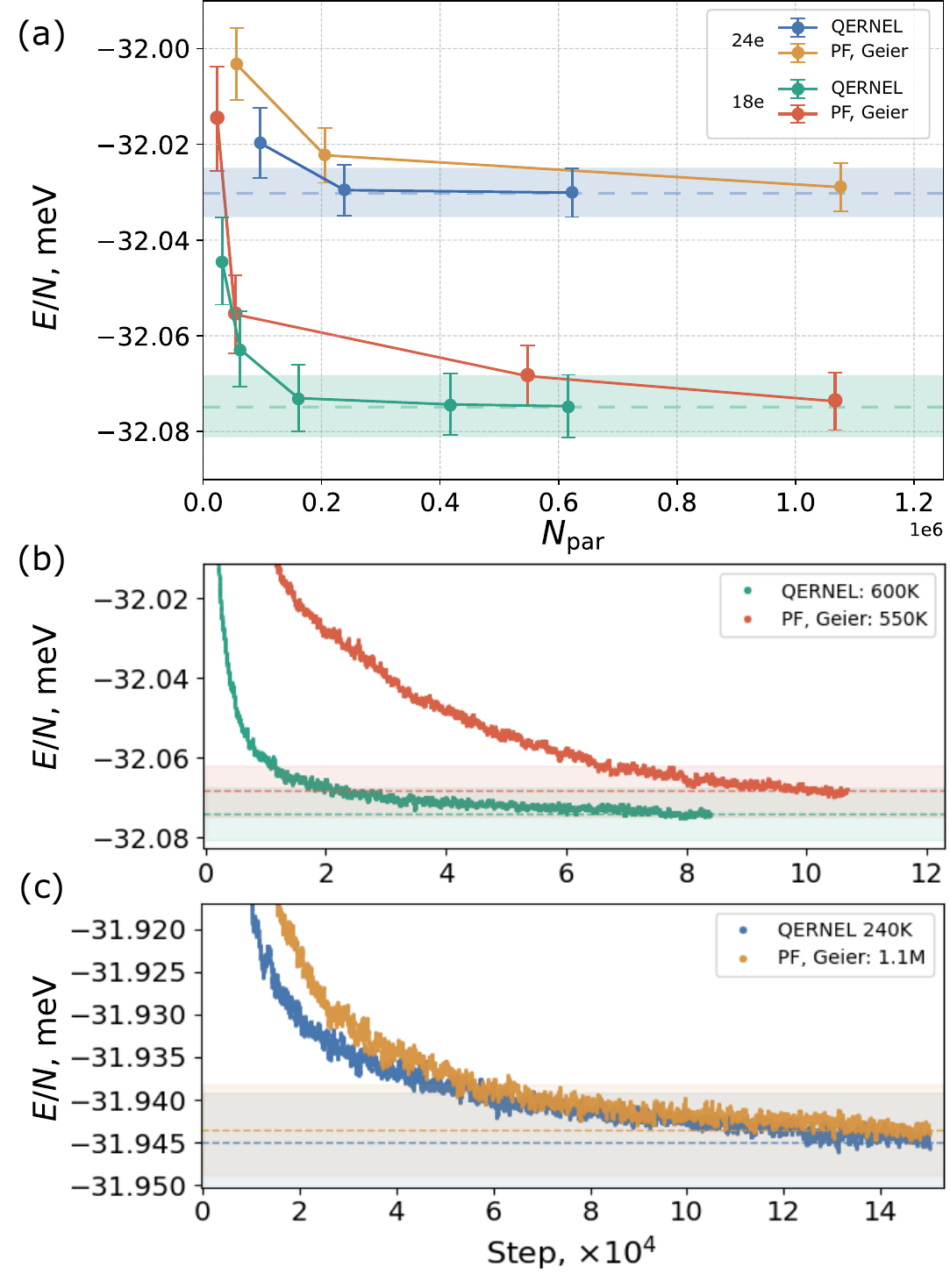}
    \caption{Benchmarking the efficiency of QERNEL. (a) final variational energy per electron versus parameter count for 18- and 24-electron systems. QERNEL reaches the same accuracy regime as the PsiFormer baseline \cite{GeierNazaryan2025} with roughly $6$--$7\times$ fewer parameters. For sake of presentation, the 24e data was shifted down by 0.08meV (b),(c) Training curves of QERNEL vs baseline.  (b) at comparable parameter count, QERNEL converges about $6\times$ faster and with improved stability than the baseline. (c) a $240$K-parameter QERNEL converges to a slightly lower energy as a $1.1$M-parameter baseline.}    \label{fig:speed}
\end{figure}

\begin{figure}
    \includegraphics[width=0.7\linewidth]{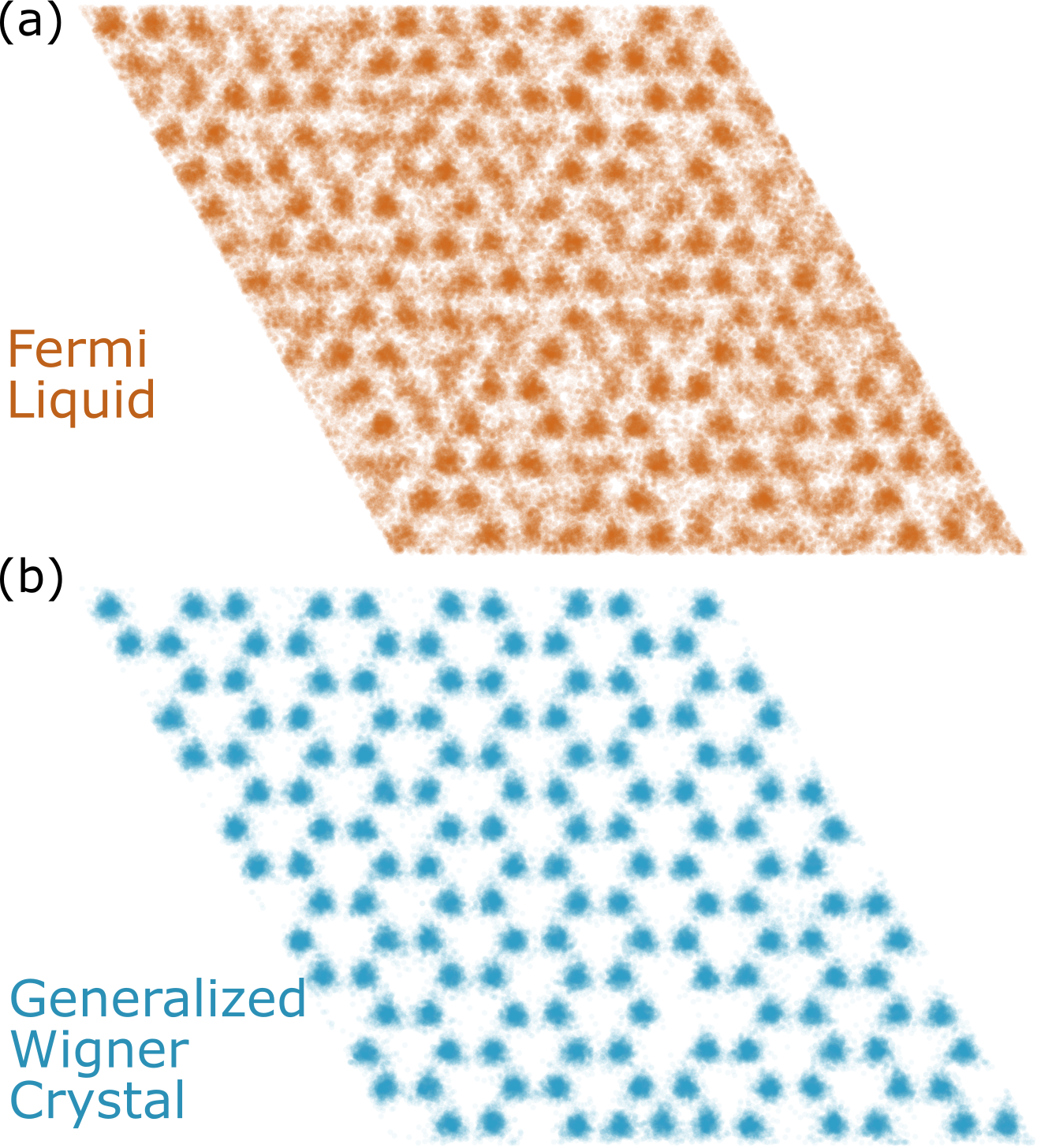}
    \caption{Inferred real space densities for a 150-electron system obtained with QERNEL in a fixed-context, non-foundational setting. (a) Fermi-liquid regime, $\lambda=0.7$. (b) Generalized Wigner-crystal regime, $\lambda=1.5$. The isolated minor defects in the density profile are a computational artifact and are not interpreted as a physical excitation.}
    \label{fig:150particles}
\end{figure}

In this work we solve these issues by introducing QERNEL: \textbf{Q}uantum \textbf{E}xpert-\textbf{R}outed \textbf{Ne}ural \textbf{L}earner. QERNEL is a large electron model designed to solve entire families of ground states within a single conditional neural wavefunction. Architecturally, it moves away from the plain Transformer (as in GPT-2) and toward the %design language 
design of modern large language models. %efficient AI models. 
Concretely, we introduce grouped-query attention (GQA) \cite{GQA2023} to capture inter-electron correlations, mixture-of-experts (MoE) \cite{MoE2017} routing to allocate capacity effectively,  feature-wise linear (FiLM-style) context modulation \cite{FiLM2018} to adapt the network across Hamiltonian parameter space, custom power-law tailed activation functions for stable training and efficient transfer learning across system size. Together, these innovations enable far fewer network parameters and much faster and more stable training. Once trained, the context conditioning allows Hamiltonian parameter sweeps to be performed by inference and short fine tuning, rather than by full retraining at every point.

We apply QERNEL to interacting electrons in semiconductor moir\'e heterobilayers with periodic boundary conditions and study the ground state across Hamiltonian parameter space within a single conditional model. Our study uncovers a sharp quantum phase transition between the Fermi liquid state and the generalized Wigner crystal. Such behavior is a stringent test of a foundational ansatz: the model must represent both phases within one shared wavefunction family and still resolve their abrupt reorganization, rather than smoothing them into a single interpolation. We show that QERNEL succeeds in doing so, demonstrating that a context-aware foundational model can capture the sharp transition that emerges in a quantum many-body phase diagram. Owing to the efficiency of the architecture, this can be done for substantially larger systems than would otherwise be accessible, reaching up to 96 electrons in the foundational setting. To further illustrate scalability beyond the conditional setting, we also perform fixed-context calculations at selected parameter values for systems of up to 150 electrons, where the two phases remain clearly distinguishable in real space.

A broader objective of this work is to make the quantum foundation model more computationally accessible to the wider condensed-matter community. QERNEL is designed not only to improve variational accuracy, but to do so on modest hardware. All calculations reported in this work were carried out on only two NVIDIA V100 GPUs, with a total of 64\,GB of GPU memory. We therefore view architectural efficiency not merely as a technical refinement, but as a practical step toward making context-aware neural many-body methods reproducible and usable by a broader community. %We plan to share the codebase upon acceptance of this work.

\textit{QERNEL ansatz.} QERNEL is a variational ansatz for a continuously parameterized family of ground states. As shown in Fig.~\ref{fig:architecture}, it maintains one feature stream per electron, updates these streams through permutation-equivariant layers, and maps them to generalized orbitals whose Slater determinant enforces fermionic antisymmetry. Its distinctive feature is that these equivariant updates are built to be explicitly economical, so that the ansatz remains expressive even when trained across a broad Hamiltonian manifold.

The main block in Fig.~\ref{fig:architecture}(a) combines grouped-query attention (GQA), mixture-of-experts (MoE), residual connections, and RMS normalization. From a physics perspective, attention captures collective many-body structure, the expert module provides flexible local nonlinear processing, and the determinant restores exact fermionic exchange antisymmetry. The internal components are shown in Fig.~\ref{fig:architecture}(b,c). In the MoE layer, a single dense perceptron is replaced by several smaller expert channels together with a router (without top-K) and an always-active shared expert, allowing specialization at reduced cost. In the attention layer, many query heads are retained while key and value heads are shared across groups, preserving rich correlation channels while reducing parameter count and memory traffic. We also replace the standard exponentially saturating activation functions by a smooth power-law-tailed counterparts to improve numerical stability.

A second essential ingredient is the explicit dependence on Hamiltonian control parameters. QERNEL conditions a single network on the external context and injects this information throughout the depth of the architecture through the FiLM modulation shown in Fig.~\ref{fig:architecture}(d). We adopt FiLM because it is extremely cheap, introducing only channel-wise shifts and rescalings of the electron stream, with a parameter overhead below roughly $5\%$ in our implementations.

QERNEL also supports efficient transfer learning across system size, so that weights trained on smaller systems provide a strong warm start for larger ones. In addition, gradient accumulation allows large effective batches to be split into microbatches and recombined during backpropagation, making it possible to train larger models and larger systems within modest GPU memory budgets.

\textit{Benchmarking performance.}  
We study interacting electrons in semiconductor moir\'e heterobilayers, where electrons reside on one semiconductor layer and experience a moir\'e potential. The Hamiltonian is expressed as 
\begin{align}
    &H = H_0 + H_{ee} \nonumber\\
    & = \sum_i \left( 
-\frac{1}{2 m^*}\nabla_i^2 + V(\mathbf{r}_i) 
\right) 
+ \frac{1}{2 \epsilon} \sum_i \sum_{i \neq j} \frac{1}{|\mathbf{r}_i - \mathbf{r}_j|}, 
\label{eq:system-hamiltonian}
\end{align}
where $V(\mathbf{r}) = -2 \lambda \cdot V_0 \sum_{j=1}^3 \cos(\mathbf{g}_j \cdot \mathbf{r} + \varphi)$ is the moir\'e potential with reciprocal lattice vectors $\mathbf{g}_j = \frac{4\pi}{\sqrt{3}a_M} (\cos \frac{2\pi j}{3}, \sin \frac{2\pi j}{3})$, moir\'e lattice constant $a_M$, and $\varphi$ controls the shape of the moir\'e potential. The parameter $\lambda$ controls the strength of the Moire potential ontop of the base $V_0 = 15\ {\rm meV}$. The physical parameters are $m^* = 0.35 m_e$,  $\varphi = \pi / 4$, a moir\'e lattice period of $a_M = 8.031 \ {\rm nm}$. In this wotk we specifically focus at $\nu=2/3$ filling per triangular potential unit cell.

A central goal of QERNEL is to increase expressivity at fixed capacity. Therefore, we benchmark it against previous transformer-based neural wavefunctions \cite{GeierNazaryan2025} in terms of both final variational accuracy and optimization speed. Fig.~2(a) shows the achieved energies versus parameter count for 18- and 24-electron systems. In both cases, QERNEL reaches the same accuracy regime as substantially larger Psiformer baselines, with comparable energies obtained using roughly $6$--$7\times$ fewer parameters.

\begin{figure*}
    \includegraphics[width=1\linewidth]{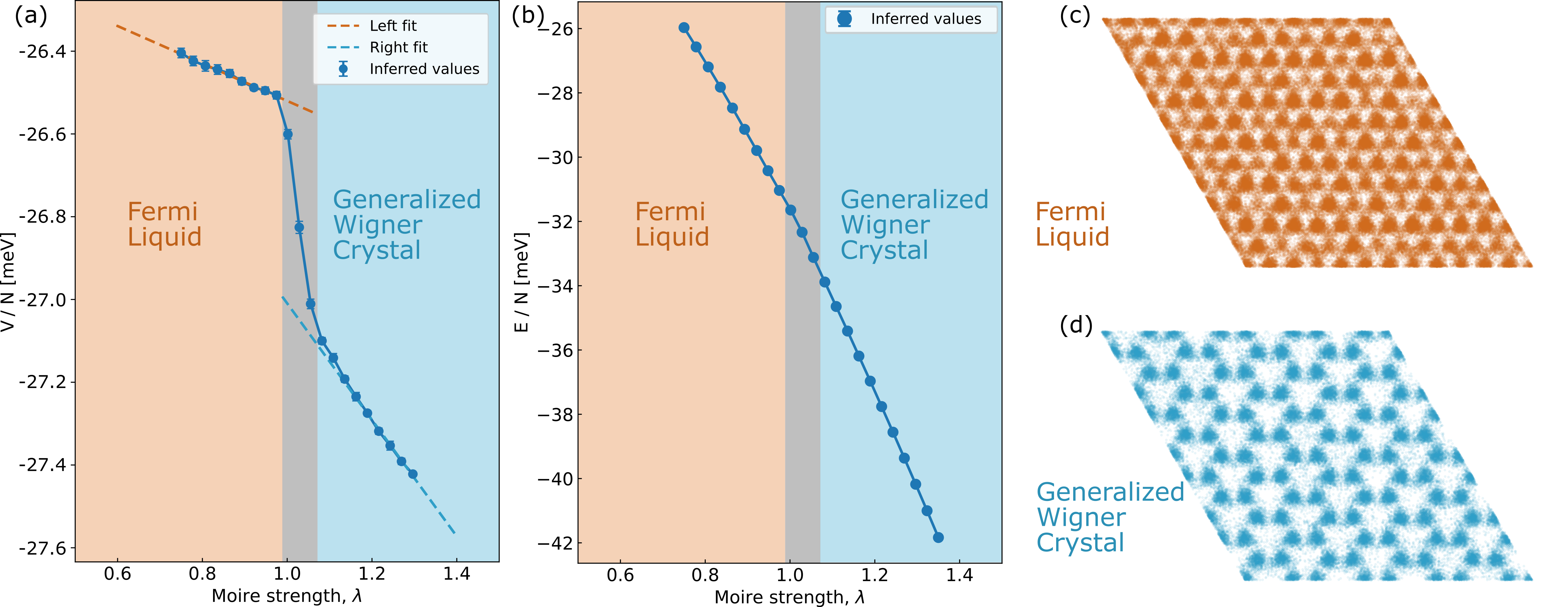}
    \caption{Inference of the foundational-model for capturing the phase transition. (a) Interaction energy per electron as a function of moir\'e strength $\lambda$, inferred from a single trained conditional QERNEL ansatz for 54-electron system. The data separate into two approximately linear branches whose extrapolated mismatch near $\lambda \approx 1.05$ indicates a sharp reorganization of the ground state. (b) Total energy per electron across the same parameter range. In contrast to the interaction energy, the total energy remains smooth and continuous through the transition region, as expected for a finite-size system. (c),(d) Inferred real-space densities on the two sides of the transition for 96-electron system. (c) The weak-moir\'e phase is a Fermi-liquid, (d) the strong-moir\'e phase has sharply localized density peaks characteristic of a generalized Wigner-crystal state.}
    % \caption{a) Interaction energy per particle as a function of moir\'e strength $\lambda$ for the 54-electron system, inferred from a single trained conditional QERNEL ansatz. The data follow distinct approximately linear trends on the two sides of the transition region. Their extrapolated mismatch near $\lambda \approx 1.05$ indicates a sharp reorganization of the ground state and is consistent with a finite-size signature of a first-order phase transition.
    % b), c) TODO
    % }
    \label{fig:transition_interaction}
\end{figure*}

Equally important is the effect on optimization. Fig.~2(b) shows that a QERNEL model with only $240\,\mathrm{K}$ parameters reaches essentially the same energy scale as a much larger $1.1\,\mathrm{M}$-parameter Psiformer. Fig.~2(c) then isolates the architectural gain at approximately matched parameter count: here QERNEL converges to its asymptotic energy about $6$ times faster in training steps, while following a smoother and more stable trajectory. The benefit of the efficient blocks is therefore twofold: they reduce the parameter budget needed for a given accuracy and accelerate optimization even when the raw parameter count is held fixed. These gains also translate into access to substantially larger systems, as we have applied our architecture in a fixed-context setting to reach 150 electrons, as illustrated in Fig.~\ref{fig:150particles}. The real-space density for 150 particles reveals the physical nature of two dinstinct phases. On the weak-moir\'e side, Fig.~\ref{fig:150particles}a), the inferred density is a Fermi-liquid phase, in which the electrons remain mostly mobile and only weakly pinned by the moir\'e landscape. On the strong-moir\'e side, by contrast, the density reorganizes into a crystal phase which breaks the translational symmetry of the underlying potential, shaping into a honeycomb lattice Fig.~\ref{fig:150particles}b).

% \section{Capturing the phase transition with a foundational model}

% \subsection{Training the model}
\textit{Training the foundational model.} 

Training a conditional wavefunction is more subtle than ordinary VMC, because the target distribution changes not only as the network parameters evolve, but also as the Hamiltonian itself changes with the context. If the control parameter is varied too rapidly, the Markov chains cannot re-equilibrate, leading to noisy energies and unreliable gradients. We therefore use a blockwise context schedule rather than resampling the context at every optimization step.

Specifically, every $K_\lambda \sim 500$ optimization steps we resample a set of $N_\lambda$ context values $\lambda$ from the target interval, partition the walkers into context blocks, and keep each block fixed for the next $K_\lambda$ steps. The model is then optimized with the averaged conditional loss
\begin{equation}
    L = \frac{1}{N_\lambda}\sum_{\lambda} \frac{\langle \psi_\theta(x, \lambda)|\hat{H}(\lambda)| \psi_\theta(x, \lambda)\rangle}{\langle \psi_\theta(x, \lambda)| \psi_\theta(x, \lambda)\rangle}.
\end{equation}
This allows a single update to probe several points in parameter space while still giving the walkers attached to each context enough time to track the corresponding equilibrium distribution. Whenever the context blocks are refreshed, we perform a short burn-in before continuing the training.

A practical tradeoff is how many context values to include in parallel. Too few give cleaner gradients but poor coverage of the Hamiltonian family; too many improve coverage but leave each context statistically noisy. We therefore use a two-stage schedule: first train with a small number of context blocks, typically $N_\lambda=4$, to separate the phases coarsely, and then increase to, e.g., $N_\lambda=16$ to refine the transition region. After training, we sweep through context values one at a time, re-equilibrate the walkers, and measure observables without further optimization. Parameter sweeps are thus performed by inference with a single learned wavefunction family rather than by retraining separate ans\"atze at each point.

% \section{Capturing phase transition with Foundational model}

% \subsection{Training the model}

% Every $K$ optimization steps (typically $K\sim200-500$), we resample several values (typically $\sim 4-8$) of the context $c$ from a prescribed range (e.g., $V_0 \in\left[V_{\text {min}}, V_{\text {max}}\right]$) and then hold $c$ fixed for the next $K$ steps while training the model under Hamiltonian $H(c)$. Keeping the context fixed in blocks gives the MCMC sampler time to re-equilibrate from the previous distribution. To avoid autocorrelation, we burn-in the MC walkers upon resampling $c$. This schedule trains a single conditional model $(x, c) \mapsto \psi_\theta(x ; c)$. We log per-context energies and observables throughout training and, at test time, evaluate interpolation/extrapolation across $\boldsymbol{c}$ with a forward pass and short sampling, without retraining.

\textit{Results.}
We now turn to the central question of this work: whether a single conditional ansatz can faithfully represent qualitatively distinct many-body states across Hamiltonian space. After training QERNEL as a foundational model over the moir\'e-strength parameter $\lambda$, we infer observables across the phase diagram without retraining the network at each point. In this way, one and the same learned wavefunction family is used to probe both sides of the transition region.

A particularly clear signature appears in the interaction energy. Figure~\ref{fig:transition_interaction}a shows the inferred interaction energy per electron for the 54-electron system as a function of moir\'e strength. Away from the transition region, the data are well described by two approximately linear branches, one on the weak-moir\'e side and one on the strong-moir\'e side. These branches do not connect smoothly near $\lambda \approx 1.05$, but instead exhibit a pronounced offset, corresponding to a sharp drop in the interaction energy. The mismatch of the left and right fits indicates that the underlying change is sharp rather than a gradual crossover. Importantly, this structure is recovered from a single context-conditioned model, showing that the foundational ansatz resolves two distinct many-body regimes within one shared representation.

At the same time, the total energy remains smooth through the same parameter range, as shown in Fig.~\ref{fig:transition_interaction}b. This is an important consistency check: while a transition is naturally revealed in individual energy components, the variational ground-state energy itself should remain continuous in a finite-size system. The transition is revealed by the real-space density profile: the Fermi-liquid phase in the weak-moir\'e regime to the generalized Wigner-crystal state for strong regime. The abrupt reconstruction of the density mirrors the discontinuous behavior seen in the interaction energy and provides direct real-space evidence for the transition between the two competing phases. The abrupt reconstruction of the density mirrors the discontinuous behavior seen in the interaction energy and provides direct real-space evidence for the transition between the two competing phases.

% Taken together, these results show that QERNEL captures both phases within a single shared conditional ansatz and resolves the sharp boundary between them. This is precisely the nontrivial requirement for a foundational model of quantum matter: it must not only interpolate across nearby Hamiltonians, but also remain expressive enough to represent distinct many-body states and the abrupt reconstruction that separates them. Owing to the efficiency of the architecture, this can be demonstrated for systems as large as 96 electrons.

\textit{Summary.}
We introduced QERNEL, a foundational neural variational ansatz that learns a family of many-body ground states within a single model. By combining efficient permutation-equivariant blocks with explicit Hamiltonian conditioning, QERNEL achieves substantially higher expressivity per unit compute than previous transformer-based neural wavefunctions, reaching comparable accuracy with several-fold fewer parameters and markedly faster, more stable training. Applied to a system of up to 150 interacting electrons in semiconductor moir\'e heterobilayers, the model captures a sharp phase transition within one shared conditional representation: the interaction energy exhibits a clear branch mismatch across the transition region, while the real-space density rapidly reorganizes from a Fermi-liquid regime into a Wigner-crystal-like state. These results show that foundational neural wavefunctions can learn not only smooth parameter dependence, but also the sharp structure of a quantum many-body phase diagram.

\textit{Acknowledgment.} We acknowledge the MIT SuperCloud and Lincoln Laboratory Supercomputing Center for providing computing resources that have contributed to the research results reported within this paper. 

\bibliography{biblio.bib}

% \section{Supplement}

\end{document}